\documentclass[pra,twocolumn]{revtex4-1}   
\usepackage{amssymb}
\usepackage{bbm,graphicx}   

\def\ave#1{\langle #1 \rangle}
\def\ii{{\rm i}}

\def\bmu{\bar{\mu}}
\def\tr#1{{\rm tr}{\,#1}}
\def\Gr{\Gamma_{\rm R}}
\def\Gl{\Gamma_{\rm L}}

\def\tit#1{{\em #1},}
\def\etal#1{ {\em et al.}}

\begin{document}

\title{Entanglement in stationary nonequilibrium states at high energies}

\author{Marko \v Znidari\v c}
\address{Faculty of Mathematics and Physics, University of Ljubljana, Ljubljana, Slovenia}

\begin{abstract}
In recent years it has been found that quantum systems can posses entanglement in equilibrium thermal states provided temperature is low enough. In the present work we explore a possibility of having entanglement in nonequilibrium stationary states. We show analytically that in a simple one-dimensional spin chain there is entanglement even at the highest attainable energies, that is, starting from an equilibrium state at infinite temperature, a sufficiently strong driving can induce entanglement, even in the thermodynamic limit. We also show that dissipative dephasing, on the other hand, destroys entanglement. 
\end{abstract}

\date{\today}

\maketitle

\section{Introduction}
It has been realized that entanglement is rather ubiquitous at sufficiently low temperatures. Ground states of interacting quantum systems typically posses entanglement between different parts of a system. Similarly, when a system is at thermal equilibrium at sufficiently low temperature entanglement is often present. This has been studied in a number of works, see e.g., Refs.~\cite{thermal} and references in~\cite{rmp}, and is called thermal entanglement~\cite{aba}.

Much less in known on the other hand about entanglement in nonequilibrium stationary states. The main difficulty is that solving a nonequilibrium system, either numerically or analytically, is much harder. Most studies of entanglement in a stationary nonequilibrium setting have so far considered only small systems of 2 or 3 spin-$1/2$ particles~\cite{small}. It has been shown that currents can enhance entanglement~\cite{current}. In the present work we present fully analytic results for entanglement in a stationary state of a nonequilibrium system of arbitrary size. The main questions we study are the dependence of entanglement in a nonequilibrium stationary state (NESS) on the coupling strength to the bath, the driving strength measuring how far we are from equilibrium, and the dependence on system size. The Hamiltonian part of the model we consider is a one-dimensional quantum spin $1/2$ chain with XX-type nearest-neighbor coupling,
\begin{equation}
H=\sum_{j=1}^{n-1} \sigma_j^{\rm x}\sigma_{j+1}^{\rm x}+\sigma_j^{\rm y}\sigma_{j+1}^{\rm y}.
\label{eq:XX}
\end{equation}
The XX model in its unitary setting is rather simple; it can be mapped to a system of noninteracting spinless fermions and can be easily diagonalized. A nonequilibrium variant, in which we couple the system to reservoirs at chain ends, can also be solved analytically, even in the presence of dissipative dephasing. Nonequilibrium dynamics of the XX model with or without dephasing~\cite{Karevski:09,JSTAT10,JPA10,PRE11,temme:09,eisler11} is rather rich, with transport varying from ballistic to diffusive.

\begin{figure}[h]
\centerline{\includegraphics[width=0.4\textwidth]{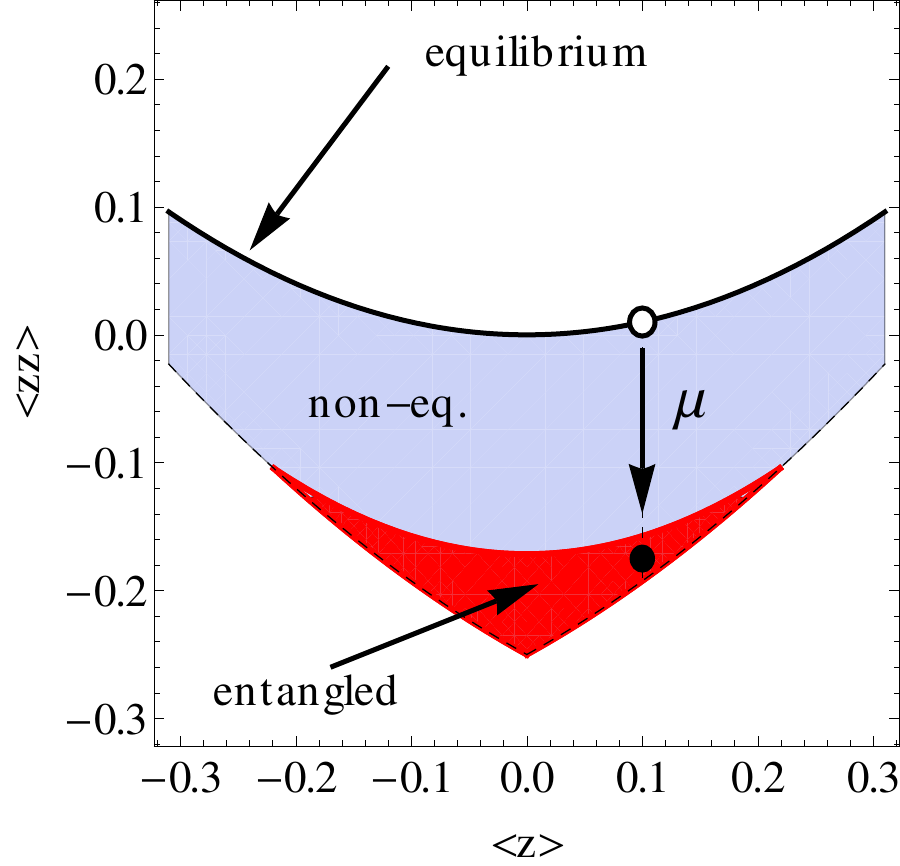}}
\caption{(Color online) Expectation values of $\ave{\sigma_j^{\rm z}}\equiv \ave{z}$ and $\ave{\sigma_j^{\rm z}\sigma_{j+1}^{\rm z}}\equiv \ave{zz}$ in the bulk of a NESS for the XX model without dephasing. The thick top curve is equilibrium states (all at infinite temperature), the shaded region is NESS states which can be reached by nonzero driving $\mu$. As one increases $\mu$ the NESS state moves downward, eventually reaching a region of NESSs having entanglement (dark red region with full circle). For details see text.}
\label{fig:smile}
\end{figure}
For the XX model without dephasing, which shows ballistic transport and is coherent in the bulk, our findings are summarized in Figure~\ref{fig:smile}. As one turns on the driving $\mu$ one moves away from an equilibrium state at infinite temperature (indicated by an empty circle in Fig.~\ref{fig:smile}), eventually reaching NESS in which one has entanglement between nearest neighbor spins, independent of system size. By a sufficiently strong driving one can induce entanglement even in an infinite temperature equilibrium state. A nonequilibrium setting therefore enlarges the region of parameters for which entanglement is present. In the presence of dephasing though, situation changes. Entanglement is present only if dephasing is sufficiently small, or if the system is small enough. In this diffusive regime there is no entanglement in the thermodynamic limit.

\section{Thermal entanglement}
Before going to nonequilibrium properties let us first remind ourselves of the entanglement in equilibrium state. Results in this section are not new and will serve as a reference point against which we will compare NESS results. For a more detailed study of entanglement in the XY model see, for instance, Ref.~\cite{Maziero10}. As a separability criterion we will always use the minimal eigenvalue of the partially transposed two-spin reduced density matrix $\rho_{j,j+1}^{\rm PT}$, where transposition is done with respect to one spin. Its negativity is a necessary and sufficient condition for entanglement of two spin-$1/2$ particles~\cite{PPT} which is the type of entanglement that we study throughout the paper.

\begin{figure}
\centerline{\includegraphics[width=0.47\textwidth]{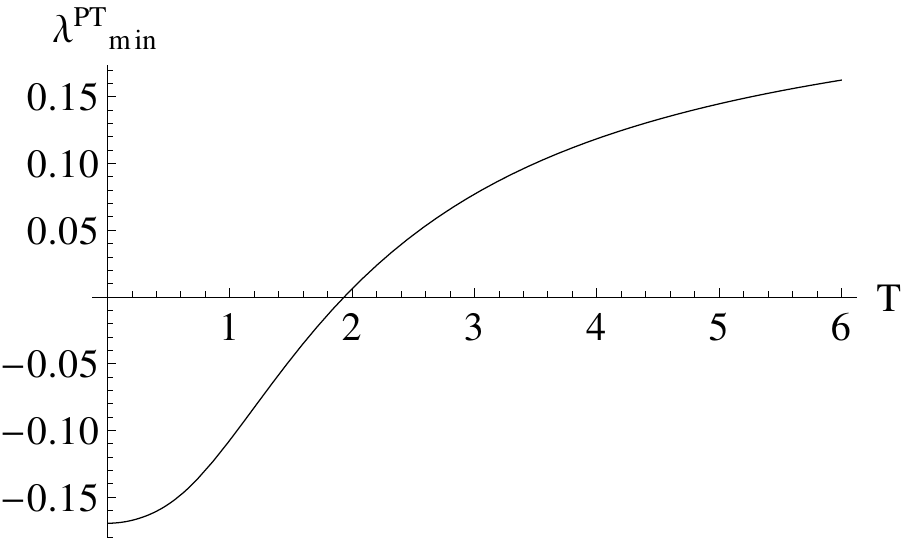}}
\caption{Minimal eigenvalue of $\rho_{j,j+1}^{\rm PT}$ (\ref{eq:Glmin}) for a grand-canonical state of XX model at $\phi=0$ (\ref{eq:Gexpect}). A two-spin reduced density matrix $\rho_{j,j+1}$ is entangled for $T< 1.94$.}
\label{fig:lminT}
\end{figure}

We shall consider the grand-canonical state $\rho_{G}$ at a certain inverse temperature $\beta=1/T$ and chemical potential $\phi$,
\begin{equation}
\rho_{\rm G}=\frac{\exp{(-\beta\, H-\phi\, M )}}{Z},\quad Z=\tr{({\rm e}^{(-\beta H-\phi M )})},
\label{eq:G}
\end{equation}
where $M=\sum_{j=1}^n \sigma_j^{\rm z}$ is total magnetization. To calculate the reduced density matrix of two nearest neighbor spins in an infinite grand-canonical chain $\rho_{\rm G}$ we need all two spin grand-canonical expectation values $\ave{\sigma_j^{\alpha} \sigma_{j+1}^{\alpha'}}_{\rm G}$, where indices $\alpha, \alpha'$ label three Pauli matrices and an identity, and $\ave{A}_{\rm G}\equiv \tr{(\rho_{\rm G}A)}$. Due to the symmetry of the XX model the only nonzero terms are $\ave{\sigma_j^{\rm z}}_{\rm G}$ and $\ave{\sigma_j^{\rm x} \sigma_{j+1}^{\rm x}}_{\rm G}=\ave{\sigma_j^{\rm y} \sigma_{j+1}^{\rm y}}_{\rm G}=\ave{H}_{\rm G}/(2n)$ and $\ave{\sigma_j^{\rm z} \sigma_{j+1}^{\rm z}}_{\rm G}$. For an infinite chain the energy density and magnetization can be found in~\cite{Katsura62}, whereas correlations are in Ref.~\cite{Barouch71}. In an infinite chain the relation $\ave{\sigma_j^{\rm z} \sigma_{j+1}^{\rm z}}_{\rm G}=\ave{\sigma_j^{\rm z}}_{\rm G}^2-\ave{\sigma_j^{\rm x}\sigma_{j+1}^{\rm x}}_{\rm G}^2$ holds~\cite{Barouch71}. In the thermodynamic limit $n \to \infty$ we have average magnetization and energy density
\begin{eqnarray}
\label{eq:Gexpect}
\ave{\sigma_j^{\rm z}}_{\rm G}&=& \frac{1}{\pi} \int_0^\pi{\tanh{\left( \phi-2\beta \cos{k}\right)}\, {\rm d}k},\\
\ave{\sigma_j^{\rm x}\sigma_{j+1}^{\rm x}}_{\rm G}&=& \frac{1}{\pi} \int_0^\pi{\cos{(k)}\tanh{\left( \phi-2\beta \cos{k}\right)}\, {\rm d}k}. \nonumber
\end{eqnarray}
Using these one can easily write a reduced density matrix of two nearest-neighbor spins, $\rho_{j,j+1}=\frac{1}{4}[\mathbbm{1}+\ave{\sigma_j^{\rm z}}_{\rm G}(\sigma_j^{\rm z}+\sigma_{j+1}^{\rm z})+\ave{\sigma_j^{\rm x}\sigma_{j+1}^{\rm x}}_{\rm G}(\sigma_j^{\rm x}\sigma_{j+1}^{\rm x}+\sigma_j^{\rm y}\sigma_{j+1}^{\rm y})+\ave{\sigma_j^{\rm z}\sigma_{j+1}^{\rm z}}_{\rm G}\, \sigma_j^{\rm z}\sigma_{j+1}^{\rm z} ]$, from which one can express the minimal eigenvalue of the partially transposed reduced density matrix $\rho_{j,j+1}^{\rm PT}$. The result is
\begin{equation}
\lambda^{\rm PT}_{\rm min}=\frac{1}{4}\left(1-\ave{\sigma_j^{\rm x}\sigma_{j+1}^{\rm x}}_{\rm G}^2+\ave{\sigma_j^{\rm z}}_{\rm G}^2-2\sqrt{w} \right),
\label{eq:Glmin}
\end{equation}
where $w=\ave{\sigma_j^{\rm x}\sigma_{j+1}^{\rm x}}_{\rm G}^2+\ave{\sigma_j^{\rm z}}_{\rm G}^2$. Transition from an entangled state to a separable state happens when $\lambda_{\rm min}^{\rm PT}=0$, which gives a condition $(\ave{\sigma_j^{\rm x}\sigma_{j+1}^{\rm x}}_{\rm G}+\sqrt{2})^2=1+\ave{\sigma_j^{\rm z}}_{\rm G}^2$. Numerically solving this equation for critical $\beta$ at $\phi=0$ we get $\beta_{\rm c} \approx 0.5162$, or temperature $T_{\rm c} \approx 1.94$. This means that below this temperature nearest-neighbor spins in a grand-canonical state of the XX model are entangled, see Fig.~\ref{fig:lminT}. Critical $\beta$ varies with $\phi$ only very slightly, for instance, at $\phi=4$ it is $\beta_{\rm c} \approx 0.5146$.

\begin{figure}
\centerline{\includegraphics[width=0.48\textwidth]{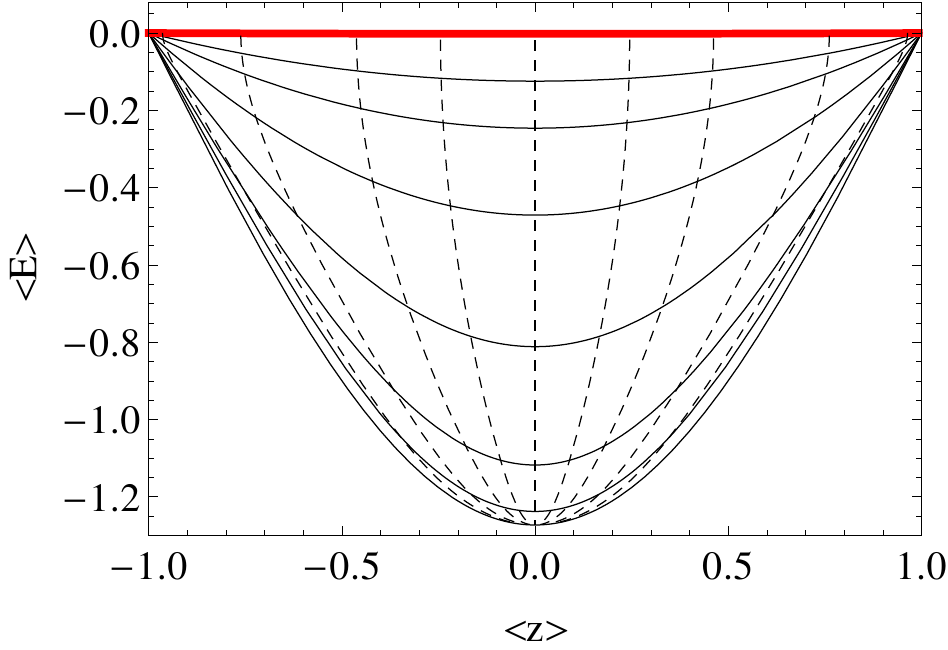}}
\caption{(Color online) Isocurves of constant $\beta$ (solid curves) and $\phi$ (dashed curves) in a grand-canonical state $\rho_{\rm G}$ (\ref{eq:G}). Expectation values of energy density $\ave{E}\equiv 2\ave{\sigma_j^{\rm x}\sigma_{j+1}^{\rm x}}_{\rm G}$ and $\ave{z}\equiv \ave{\sigma_j^{\rm z}}_{\rm G}$ (\ref{eq:Gexpect}) in a two-spin nearest neighbor reduced density matrix $\rho_{j,j+1}$ uniquely determine $\beta$ and $\phi$. Iso-temperature curves are at $T=1/\beta=0, 0.5, 1, 2, 4, 8, \infty$ (bottom to top), while iso-$\phi$ curves are at $\phi=0,\pm 0.25, \pm 0.5, \pm 1, \pm 2$ (center to right/left). Thick (red) line is expectation values for NESS states studied in the present work.}
\label{fig:thermalexpect}
\end{figure}

Note that the expectation values of magnetization and energy density uniquely determine a two-spin nearest-neighbor reduced density matrix if the whole system is in a grand-canonical state. The relation between these two expectation values and temperature $T$ and chemical potential $\phi$ can be inferred from Fig.~\ref{fig:thermalexpect}. As we shall see, all NESS states studied in the present work have energy density equal to zero, and therefore lie on the thick (red) line in Fig.~\ref{fig:thermalexpect}, that is, they have the same expectation values of energy and magnetization as equilibrium grand-canonical states at infinite temperature, $\beta=0$. Note, however, that such NESS states are of course not grand-canonical as other expectation values, for instance the current or $\ave{\sigma_j^{\rm z}\sigma_{j+1}^{\rm z}}_{\rm G}$ do not have the same expectation values as in the grand-canonical state.

\section{Nonequilibrium XX model}
\subsection{Setting}
The nonequilibrium dynamics of the XX model will be described by the Lindblad master equation~\cite{lindblad},
\begin{equation}
\frac{{\rm d}}{{\rm d}t}{\rho}=\ii [ \rho,H ]+ {\cal L}^{\rm bath}_{\rm L}(\rho)+{\cal L}^{\rm bath}_{\rm R}(\rho)+{\cal L}^{\rm deph}(\rho).
\label{eq:Lin}
\end{equation}
Each of the three dissipative terms is expressed in terms of Lindblad operators $L_k$ as
\begin{equation}
{\cal L}^{\rm dis}(\rho)=\sum_k \left( [ L_k \rho,L_k^\dagger ]+[ L_k,\rho L_k^{\dagger} ] \right).
\end{equation}
Dephasing is described by $n$ Lindblad operators, each acting only on the $j$-th spin, and being
\begin{equation}
L^{\rm deph}_j=\sqrt{\frac{\gamma}{2}}\sigma^{\rm z}_j.
\end{equation}
The bath dissipator ${\cal L}^{\rm bath}_{\rm L}$ acts only on the 1st spin while ${\cal L}^{\rm bath}_{\rm R}$ acts on the last spin. Each involves two Lindblad operators, on the left end
\begin{equation}
L^{\rm L}_1=\sqrt{\Gl(1-\frac{\mu}{2}+\bar{\mu})}\,\sigma^+_1,\quad L^{\rm L}_2=\sqrt{\Gl(1+\frac{\mu}{2}-\bar{\mu})}\, \sigma^-_1,
\label{eq:Lbath}
\end{equation}
while on the right end we have
\begin{equation}
L^{\rm R}_1=\sqrt{\Gr(1+\frac{\mu}{2}+\bar{\mu})}\,\sigma^+_n,\quad L^{\rm R}_2=\sqrt{\Gr(1-\frac{\mu}{2}-\bar{\mu})}\, \sigma^-_n,
\end{equation}
where $\sigma^\pm_j=(\sigma^{\rm x}_j \pm {\rm i}\, \sigma^{\rm y}_j)/2$. Relevant parameters are the two coupling strengths to the baths $\Gamma_{\rm L,R}$, driving strength $\mu$ that dictates the magnetization difference between chain ends, the average driving $\bar{\mu}$ and the dephasing strength $\gamma$. Because driving coefficients have to be real we must have $\Gamma_{\rm L,R} \ge 0$  as well as (for all four sign combinations)
\begin{equation}
1\pm \frac{\mu}{2}\pm \bmu \ge 0.
\end{equation}
In order for this to hold $\mu$ and $\bmu$ must lie inside a rhombus with corners $(0,\pm 1)$ and $(\pm 2,0)$ in a ``$\mu-\bmu$'' plane. If instead of $\mu$ and $\bmu$ we define the two parameters 
\begin{equation}
\frac{\mu}{2}+\bmu \equiv d,\qquad \frac{\mu}{2}-\bmu \equiv c,
\label{eq:cd}
\end{equation}
i.e., $\mu=c+d, \bmu=(d-c)/2$, definition range is a simple square, $c \in [-1,1]$ and $d\in [-1,1]$. The XX model with or without dephasing has been studied in Refs.~\cite{Karevski:09,temme:09,JSTAT10,JPA10,eisler11,PRE11}. Our analytical calculations rely heavily on the exact NESS solution found in Refs.~\cite{JSTAT10,JPA10,PRE11}.

\subsection{Equilibrium, $\mu=0$}

The equilibrium situation is obtained for $\mu=0$ (other parameters can be arbitrary). The NESS state is, in this case, separable and equal to~\cite{PRE11},
\begin{equation}
\rho_{\rm NESS}=\frac{1}{2^n}\prod_{j=1}^n(\mathbbm{1}+\bmu\, \sigma_j^{\rm z}).
\label{eq:eq}
\end{equation}
This is exactly the grand-canonical state (\ref{eq:G}) with $\beta=0$ and $\tanh{\phi}=-\bmu$. All equilibrium states, from which NESS will be reached for nonzero $\mu$, studied in the present work therefore are at infinite temperature. These are indicated by a thick (red) line in Fig.~\ref{fig:thermalexpect}. 

\subsection{Nonequilibrium and no dephasing, $\gamma=0$}
We first focus on the case with $\gamma=0$ because the main entanglement features are already present in this simpler XX model without dephasing. Because everything is independent of the system size $n$ the results presented in this subsection are valid for any $n$.

Expectation values in the NESS that we need for a two-spin reduced density matrix are given in Ref.~\cite{JPA10},
\begin{eqnarray}
-\ave{\sigma_j^{\rm x} \sigma_{j+1}^{\rm y}} &\equiv& t = \mu \frac{\Gamma_{\rm L} \Gamma_{\rm R}}{(1+\Gamma_{\rm L} \Gamma_{\rm R})(\Gamma_{\rm L}+\Gamma_{\rm R})}, \\
\ave{\sigma_1^{\rm z}}&\equiv& a_1 =\bmu-\frac{\mu}{2}\, \frac{(\Gamma_{\rm L}-\Gamma_{\rm R})+\Gamma_{\rm L}\Gamma_{\rm R}(\Gamma_{\rm L}+\Gamma_{\rm R})}{(1+\Gamma_{\rm L} \Gamma_{\rm R})(\Gamma_{\rm L}+\Gamma_{\rm R})} \nonumber \\
\ave{\sigma_{2,\ldots,n-1}^{\rm z}}&\equiv& a=\bmu-\frac{\mu}{2}\, \frac{(\Gamma_{\rm L}-\Gamma_{\rm R})(1-\Gamma_{\rm L} \Gamma_{\rm R})}{(1+\Gamma_{\rm L} \Gamma_{\rm R})(\Gamma_{\rm L}+\Gamma_{\rm R})} \nonumber \\
\ave{\sigma_n^{\rm z}}&\equiv& a_n =\bmu-\frac{\mu}{2}\, \frac{(\Gamma_{\rm L}-\Gamma_{\rm R})-\Gamma_{\rm L}\Gamma_{\rm R}(\Gamma_{\rm L}+\Gamma_{\rm R})}{(1+\Gamma_{\rm L} \Gamma_{\rm R})(\Gamma_{\rm L}+\Gamma_{\rm R})} \nonumber,
\label{eq:a}
\end{eqnarray}
as well as $\ave{\sigma_j^{\rm z}\sigma_{k}^{\rm z}}=a_j a_k -t^2 \delta_{j+1,k}$.
Ranges that these parameters can take are $t \in \mu [0,\frac{1}{4}]$, then $a_1 \in \bmu - \frac{\mu}{2}[-1,1]$, $a \in \bmu - \frac{\mu}{2}[-1,1]$, and $a_n \in \bmu - \frac{\mu}{2}[-1,1]$.

For equal coupling strengths at both ends, $\Gamma_{\rm L}=\Gamma_{\rm R}$, which we will use in all figures, the above expressions simplify to $t=\frac{\mu}{2}\frac{\Gamma}{1+\Gamma^2}, a_1=\bmu-\Gamma t, a=\bmu, a_n=\bmu+\Gamma t$.

\subsubsection{Two nearest-neighbor spins at the boundary}
The reduced density matrix for two spins at the chain end is
\begin{equation}
\rho_{12}=\frac{1}{4}\left( 
\begin{array}{cccc}
x & 0 & 0 & 0\\
0 & y & 2\ii t & 0\\
0 & -2\ii t & z & 0\\
0 & 0 & 0 & v 
\end{array}
\right) ,
\end{equation}
with $x=1+a+a_1+aa_1-t^2$, $y=1+a-a_1-aa_1+t^2$, $z=1-a+a_1-aa_1+t^2$, $v=1-a-a_1+aa_1-t^2$. Calculating $\rho_{12}^{\rm PT}$ we get only one eigenvalue that can possibly be negative and which is 
\begin{equation}
\lambda^{\rm PT}_{\rm min}=\frac{1}{4}(1+aa_1-t^2-\sqrt{(a+a_1)^2+4t^2}).
\end{equation}

\begin{figure}[h]
\centerline{\includegraphics[width=0.25\textwidth]{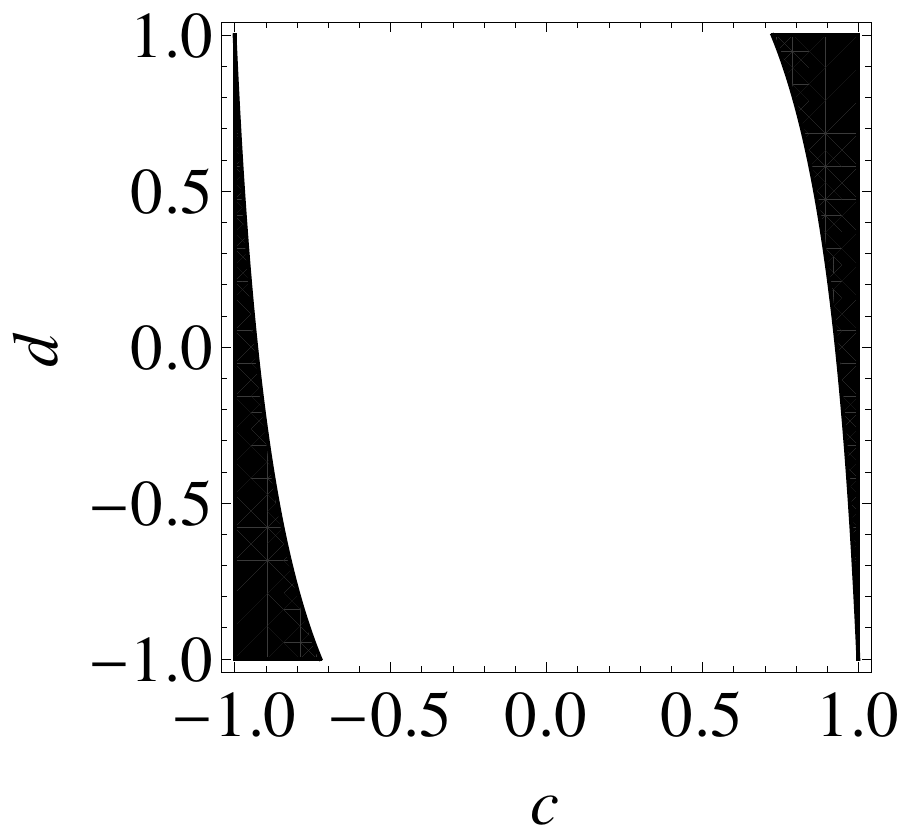}\includegraphics[width=0.25\textwidth]{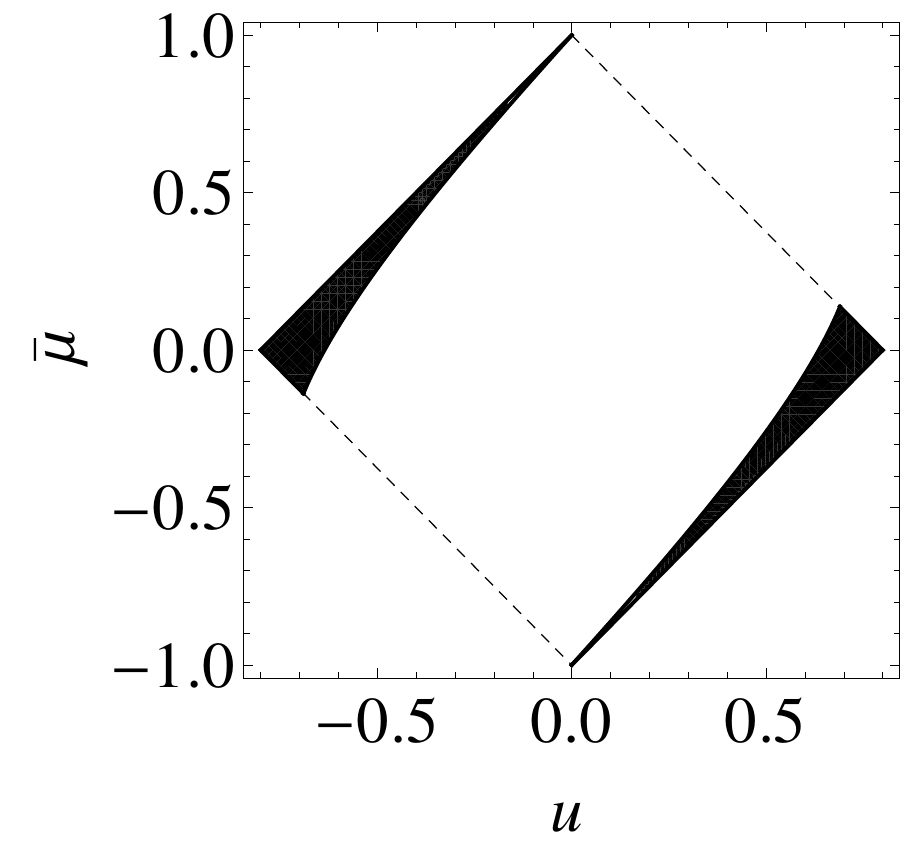}}
\caption{Parameters for which $\rho_{12}$ is entangled in the NESS (black areas). Left: values of $c$ and $d$ (\ref{eq:cd}) for which there is entanglement. Right: values of $u\equiv \Gamma t=(\ave{\sigma_n^{\rm z}}-\ave{\sigma_1^{\rm z}})/2$ and $\bmu$ with an entangled $\rho_{12}$. Right plot is just a rescaled and rotated left figure. Dashed lines in the right plot denote a region of allowed parameter values. $\Gamma_{\rm L}=\Gamma_{\rm R}=2$.}
\label{fig:bG2}
\end{figure}

\begin{figure}[h]
\centerline{\includegraphics[width=0.35\textwidth]{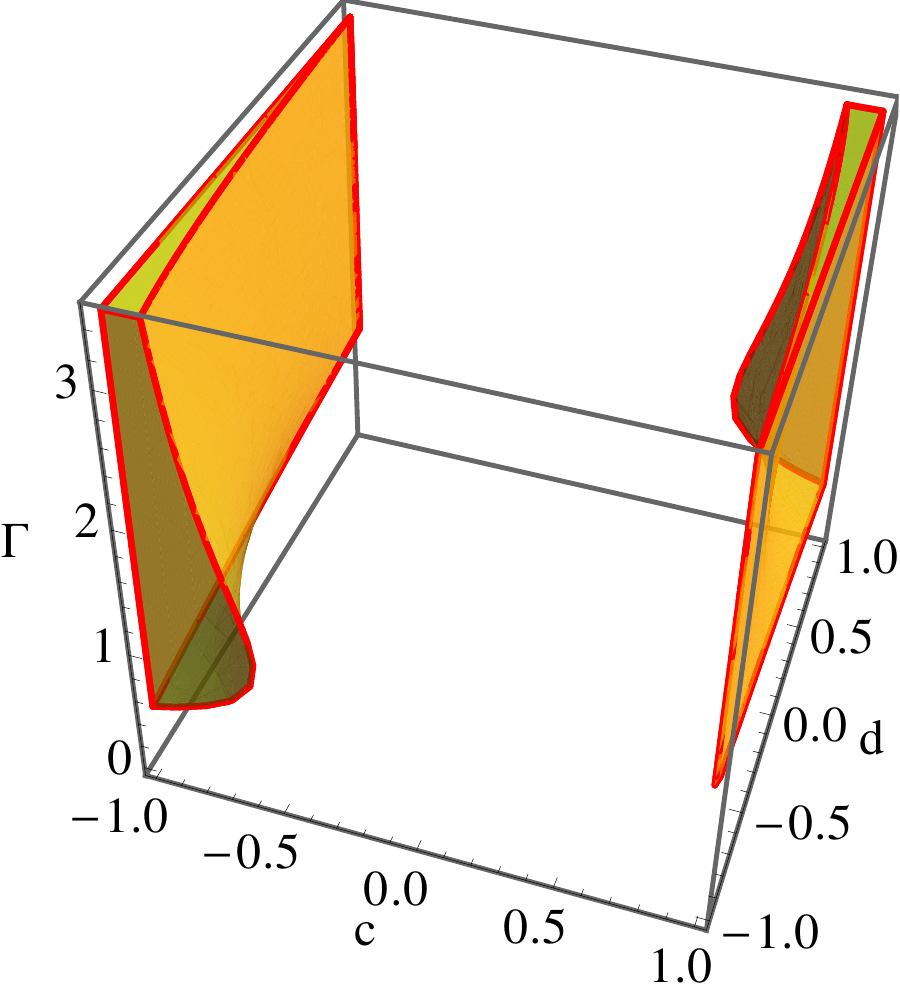}}
\caption{(Color online) Vertical axis is $\Gamma$ ($\Gamma_{\rm L}=\Gamma_{\rm R}$), horizontal axes are $c$ and $d$. Parameters within the two symmetrically placed shaded regions, existing for $\Gamma > \Gamma_{\rm min}$ (\ref{eq:Gmin}), correspond to entangled boundary two spins in the NESS. Figure~\ref{fig:bG2} shows a cross-section of this plot at $\Gamma=2$.}
\label{fig:bAll}
\end{figure}

\begin{figure}[ht!]
\centerline{\includegraphics[width=0.4\textwidth]{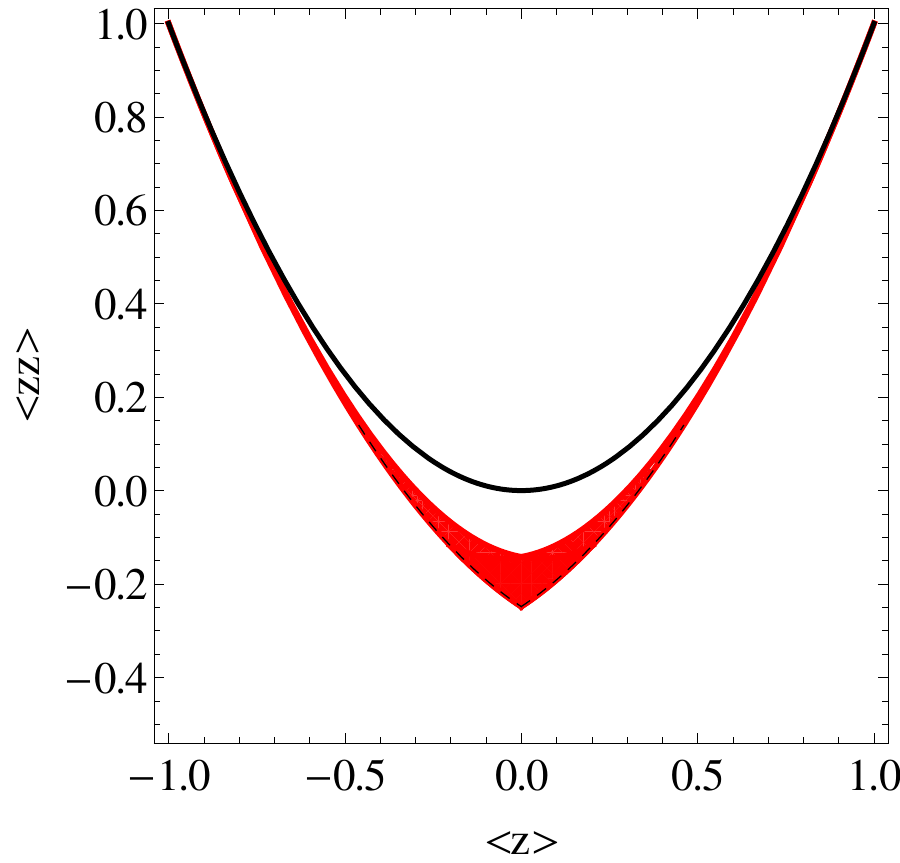}}
\caption{(Color online) Expectation values of $\ave{zz}\equiv \ave{\sigma_j^{\rm z}\sigma_{j+1}^{\rm z}}$ and $\ave{z}\equiv \ave{\sigma^{\rm z}_j}$ in the bulk of NESS lie between top solid curve and bottom dashed curve. Top solid curve corresponds to equilibrium states. As one increases $\mu$ away from zero (keeping all other parameters fixed) NESS moves vertically in the plot. For sufficiently large $\mu$ an ``entanglement smile'' region of entangled boundary two spins is reached, denoted by the dark (red) region. All is for $\Gamma=1.088$.}
\label{fig:bsmile}
\end{figure}

In order to simplify expressions, from now on in this subsection we take the same coupling at both ends, $\Gamma_{\rm L}=\Gamma_{\rm R}\equiv \Gamma$. This has no qualitatively important consequences. In Fig.~\ref{fig:bG2} we plot in black those regions of driving parameters for which $\rho_{12}$ is entangled, that is, $\lambda_{\rm min}^{\rm PT}$ is negative. 
In addition to a ``$c-d$'' plane we also show values of $u$, defined as $u\equiv\Gamma t=(\ave{\sigma_n^{\rm z}}-\ave{\sigma_1^{\rm z}})/2$ and $\bmu$, which have physical interpretation as the difference of magnetization at chain ends and the magnetization inside the chain (at all spins apart from the 1st and the last one). On the ``$c-d$'' plot the minimal $c$, i.e., the beginning of the right tongue in the left frame of Fig.~\ref{fig:bG2}, is at
\begin{equation}
c_{\rm min}=\frac{3+8\Gamma^2+4\Gamma^4-2\Gamma(1+\Gamma^2)\sqrt{8+\Gamma^2}}{1+2\Gamma^2(1+\Gamma^2)}.
\end{equation}
The functional form of the bottom boundary of this tongue, $d(c,\Gamma)$, extending in $c$ from $c_{\rm min}$ to $c=1$, is
\begin{equation}
d=1-\left(1+w-c(1+3\Gamma^2+\Gamma^4)-\sqrt{v} \right),
\end{equation}
where $w=\Gamma(1+\Gamma^2)\sqrt{8+\Gamma^2}$ and $v=(1+\Gamma^2)[(9+5c^2)\Gamma^4+(1+c^2)\Gamma^6-4(-1+cw)-2\Gamma^2(-6-2c^2+cw)]$. For $\Gamma>1$ the tongue extends all the way down to $d=-1$, that is, $d(c=1,\Gamma)=-1$. For $\Gamma<1$ it is smaller, eventually disappearing for $\Gamma<\Gamma_{\rm min}\approx 0.511$, where
\begin{equation}
\Gamma_{\rm min}=\sqrt{\frac{-3+(27-3\sqrt{78})^{1/3}+(27+3\sqrt{78})^{1/3}}{6}}.
\label{eq:Gmin}
\end{equation}
In other words, $\rho_{12}$ can be entangled only if the coupling strength is in the range $\Gamma_{\rm min}<\Gamma< \infty$. Entanglement tongues are the largest, i.e., $c_{\rm min}$ is the smallest, at $\Gamma\approx 1.08873$. Regions of $\Gamma$ and $c$ and $d$ for which $\rho_{12}$ is entangled can be seen in Fig.~\ref{fig:bAll}.
It is instructive to plot how the NESS state differs from the equilibrium grand-canonical state. As mentioned, the expectation value of the energy is for our NESS states always zero~\cite{JSTAT10,JPA10} and one can always find an equilibrium grand-canonical state having the same expectation value of energy and magnetization, see Fig.~\ref{fig:thermalexpect}. The expectation value of $\ave{\sigma_j^{\rm z}\sigma_{j+1}^{\rm z}}$ in the NESS on the other hand differs from the one in equilibrium. Therefore, we plot in Fig.~\ref{fig:bsmile} a region of allowed expectation values of $\sigma^{\rm z}_j$ and $\sigma_j^{\rm z}\sigma_{j+1}^{\rm z}$ in the bulk (away from two boundary spins). Remember that in equilibrium at infinite temperature (where $\ave{\sigma_j^{\rm x}\sigma_{j+1}^{\rm x}}_{\rm G}=0$) one has the relation $\ave{\sigma_j^{\rm z} \sigma_{j+1}^{\rm z}}_{\rm G}=\ave{\sigma_j^{\rm z}}_{\rm G}^2$ (top solid curve in Fig.~\ref{fig:bsmile}).
One can see that for a sufficiently strong driving $\mu$, keeping $\bmu$ fixed, a NESS is reached in which $\rho_{12}$ is entangled (red region). What is even more interesting, for sufficiently large $\ave{z}$, that is $\bmu$, an entangled state is reached already for very small values of $\mu$. That is, for large average magnetization $\bmu$, one can reach entanglement from an infinite temperature equilibrium state already with small driving. One can say that a small stationary perturbation of an infinite temperature equilibrium state can induce entanglement. This is quite different from thermal entanglement, which is possible only at sufficiently low temperatures.

\subsubsection{Nearest-neighbor spins in the bulk}
\begin{figure}[h]
\centerline{\includegraphics[width=0.25\textwidth]{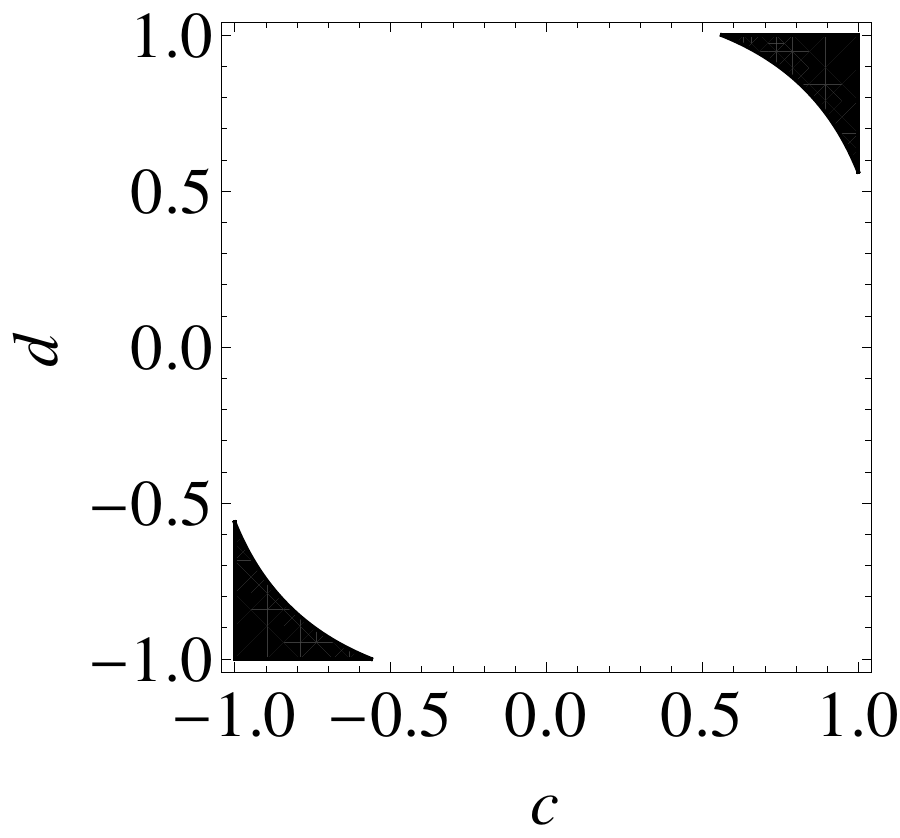}\includegraphics[width=0.25\textwidth]{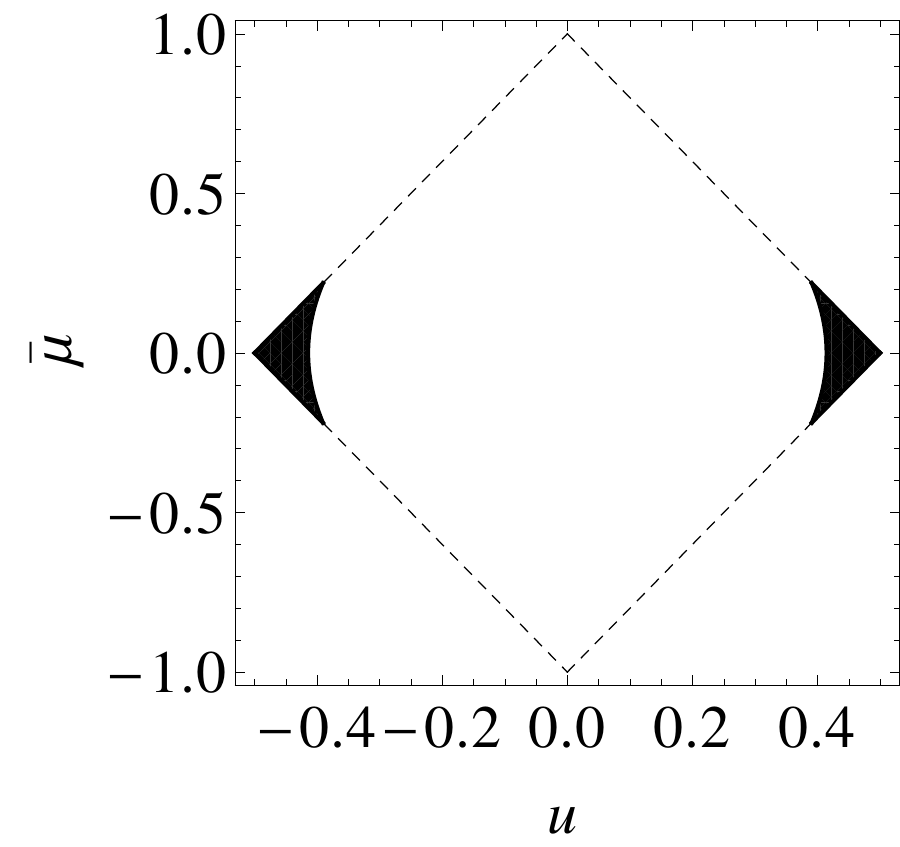}}
\caption{Regions of entanglement in $\rho_{j,j+1}$ for $\Gamma_{\rm L}=\Gamma_{\rm R}=1$ in ``$c-d$'' (left) and ``$u-\bar{\mu}$'' plane (right). There are two symmetric pockets at large $u$, that is, at large magnetization difference across the chain.}
\label{fig:bulk}
\end{figure}

\begin{figure}[h]
\centerline{\includegraphics[width=0.35\textwidth]{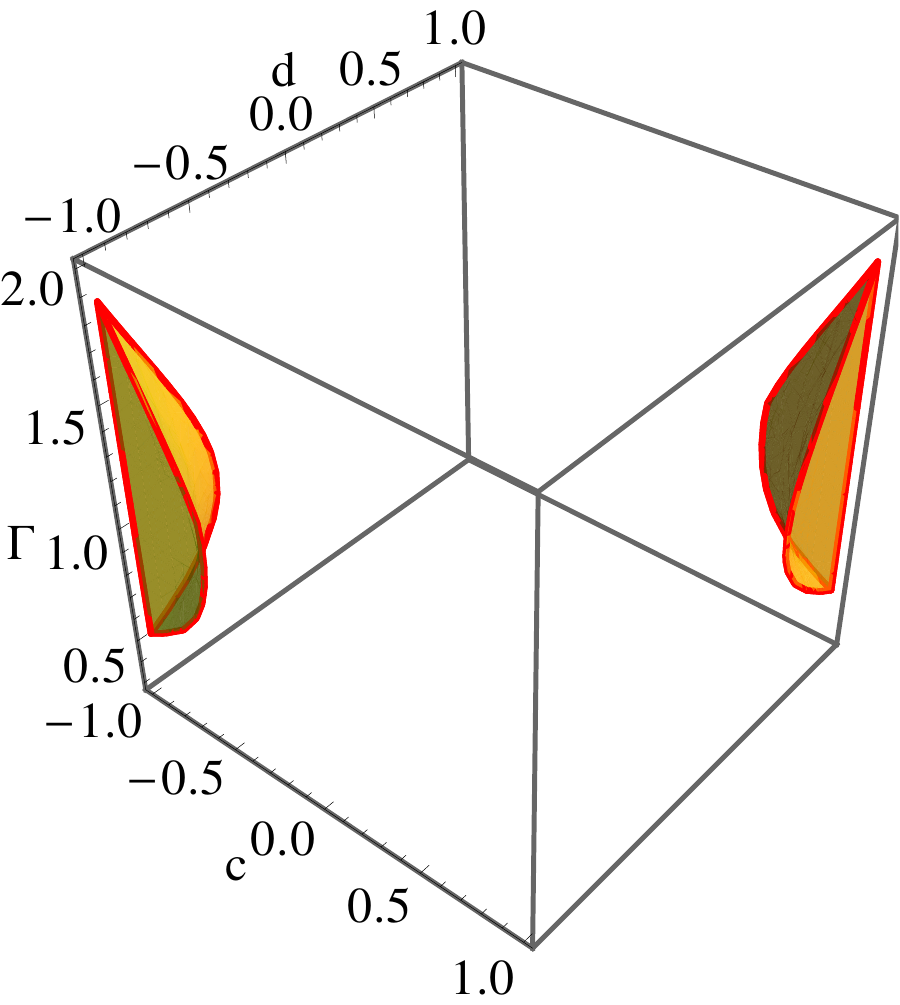}}
\caption{(Color online) Vertical axis is $\Gamma$ ($\Gamma_{\rm L}=\Gamma_{\rm R}$), horizontal are $c$ and $d$. Two pockets at intermediate couplings $0.531 < \Gamma < 1.883$ (\ref{eq:GG}) are NESSs with entangled two-spin reduced density matrix $\rho_{i,i+1}$.}
\label{fig:pockets}
\end{figure}

\begin{figure}[h]
\centerline{\includegraphics[width=0.25\textwidth]{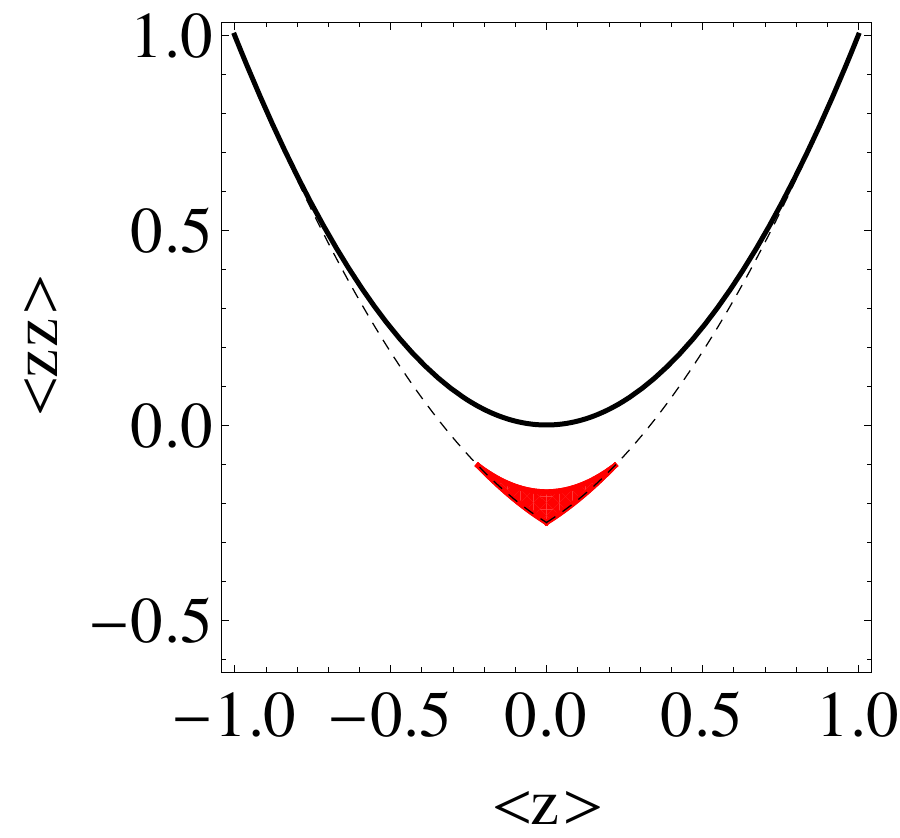}\includegraphics[width=0.25\textwidth]{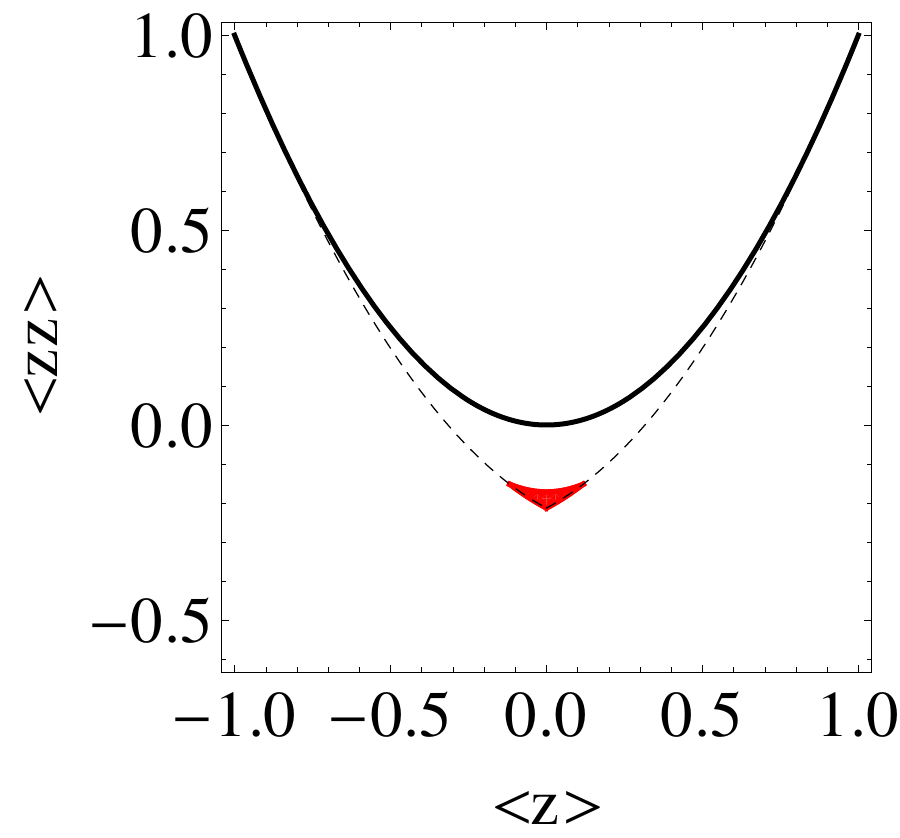}}
\caption{(Color online) Expectation values in equilibrium, $\mu=0$ (top full curve), and out of equilibrium ($\mu \neq 0$, region between dashed and full curve). Dark (red) region are NESS states with bulk nearest-neighbor entanglement. They are reached only for sufficiently strong driving $\mu$. All is for $\Gamma_{\rm L}=\Gamma_{\rm R}=\Gamma$, on the left for $\Gamma=1$, on the right for $\Gamma=1.5$. An enlarged central part of the left frame is shown in Fig.~\ref{fig:smile}.}
\label{fig:entsmile}
\end{figure}

We proceed to entanglement of two nearest-neighbor spins in the bulk of the chain. Using again expectation values (\ref{eq:a}) we get the reduced density matrix
\begin{equation}
\rho_{j,j+1}=\frac{1}{4}\left( 
\begin{array}{cccc}
A_{+} & 0 & 0 & 0\\
0 & B & 2\ii t  & 0\\
0 & -2\ii t & B & 0\\
0 & 0 & 0 & A_{-} 
\end{array}
\right) ,
\end{equation}
where $A_\pm=(1\pm a)^2-t^2$ and $B=1-a^2+t^2$. The minimal eigenvalue of the partially transposed $\rho^{\rm PT}_{j,j+1}$ is this time 
\begin{equation}
\lambda^{\rm PT}_{\rm min}=\frac{1}{4}(1+a^2-t^2-2\sqrt{a^2+t^2}).
\end{equation} 
From now on we again set $\Gamma_{\rm L}=\Gamma_{\rm R}=\Gamma$. In Fig.~\ref{fig:bulk} we can see regions of parameters for which $\rho_{j,j+1}$ is entangled. 
The main difference from the case of $\rho_{12}$ (e.g., Fig.~\ref{fig:bG2}) is that the entanglement is present in two pockets in the corners of Fig.~\ref{fig:bulk}. The lower edge of the right pocket (near corner $c=d=1$) is 
\begin{equation}
d=\frac{c(1+3\Gamma^2+\Gamma^4)+2(1+\Gamma^2)[-\Gamma\sqrt{2}+\sqrt{A}]}{1+\Gamma^2+\Gamma^4},
\end{equation}
where $A=1+\Gamma(3\Gamma+c^2\Gamma+\Gamma^3-\sqrt{8}c(1+\Gamma^2))$. It runs from $c_{\rm min}$ to $c=1$, with
\begin{equation}
c_{\rm min}=3-\frac{4\Gamma(\sqrt{2}+\Gamma(-1+\Gamma \sqrt{2}))}{1+\Gamma^2+\Gamma^4}.
\end{equation}
$c_{\rm min}$ is minimal, i.e., the pocket is the largest for $\Gamma=1$. When one increases or decreases $\Gamma$ the size of this pocket gets smaller, eventually disappearing when $c_{\rm min}=1$. This bounds the range of possible couplings to 
\begin{equation}
\frac{1}{2}(1+\sqrt{2}-\sqrt{-1+2\sqrt{2}})<\Gamma<\frac{1}{2}(1+\sqrt{2}-\sqrt{-1+2\sqrt{2}}).
\label{eq:GG}
\end{equation}
In the bulk we therefore have entanglement in NESS only in this range of $\Gamma$. This can be seen in Fig.~\ref{fig:pockets}.
In Fig.~\ref{fig:entsmile} we also plot expectation values of magnetization and $\ave{\sigma_j^{\rm z}\sigma_{j+1}^{\rm z}}$ in the bulk. 
One can see, that in contrast to entanglement in the boundary two spins, here one has entanglement only for sufficiently strong $\mu$. Note however that all these states still have zero energy density. Compare also the left plot in Fig.~\ref{fig:entsmile} with Fig.~\ref{fig:smile}, where an enlarged central section of this plot ($\Gamma=1$) is shown.

Why can entanglement at the edge of the chain be reached already for small $\mu$ while large $\mu$ is needed in the bulk? The answer lies in different magnetization values at the edge. Because the system is a ballistic transporter magnetization exhibits jump at the boundary; it is $\bmu-u$ at the 1st spin and $\bmu$ at the 2nd, causing entanglement for small $\mu$.  

Let us finally conclude the section on the XX model with a discussion of non nearest-neighbor spins. Calculating the reduced density matrix and the minimal eigenvalue of the partially transposed matrix we see that the condition for entanglement is $1-a^2 < 2\sqrt{a^2+t^4}$, in other words, $2|t|+\bmu^2 > 1$. This can never be fulfilled and there is no entanglement.

\subsection{XX with nonzero dephasing}
\begin{figure}[ht!]
\centerline{\includegraphics[width=0.45\textwidth]{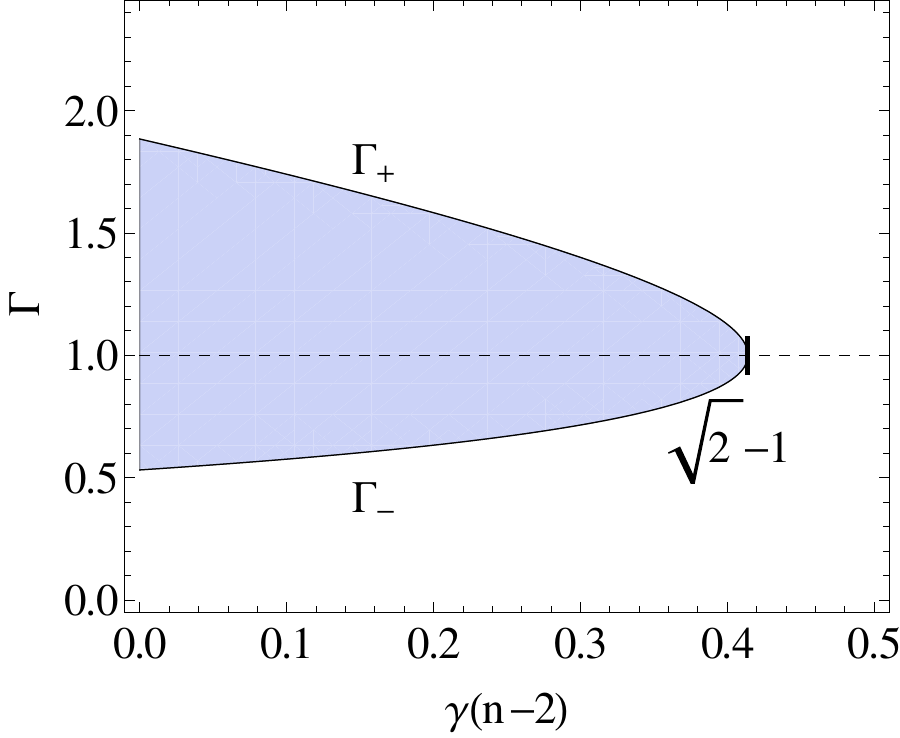}}
\caption{Shaded region is a range of coupling strengths $\Gamma$ (Eq.~\ref{eq:Gpm}) for which one can have a NESS with nearest-neighbor entanglement in the middle of the XX chain with dephasing.}
\label{fig:pocketDeph}
\end{figure}
For the XX model with dephasing NESS expectation values depend on the system size. Under nonzero driving $\mu$ the spin current is proportional to $\sim 1/n$, similarly as long-range correlations. Because dephasing diminishes off-diagonal matrix elements in the Pauli basis, it is to be expected that the entanglement will be smaller compared to a coherent XX model. This is indeed what happens. 

For simplicity we will always use $\Gamma_{\rm L}=\Gamma_{\rm R}=\Gamma$ and look at the central two spins described by the reduced density matrix $\rho_{n/2,n/2+1}$. Using the exact solution for NESS from Refs.~\cite{JSTAT10,PRE11} one can easily obtain the reduced density matrix and the minimal eigenvalue of $\rho^{\rm PT}_{n/2,n/2+1}$. Expressions for $\lambda^{\rm PT}_{\rm min}$ as well as boundaries of entangled regions are simple but long and we do not give them here. The main difference with the XX model is that the size of the entanglement pockets depends on the dephasing strength $\gamma$. There are two symmetrically placed pockets, similarly as in the case of no dephasing. Regions of entangled NESS disappear if $\gamma$ is larger than the critical $\gamma_{\rm c}$,
\begin{equation}
\gamma_{\rm c}=\frac{\sqrt{2}+1-\Gamma-1/\Gamma}{n-2}.
\label{eq:gammac}
\end{equation}
Also, regions of entangled NESS states are present only if the coupling strength $\Gamma$ satisfies $\Gamma_- < \Gamma < \Gamma_+$, where
\begin{equation}
\Gamma_\pm = \frac{1}{2}\left( w-\gamma(n-2) \pm \sqrt{(w-\gamma(n-2))^2-4} \right),
\label{eq:Gpm}
\end{equation}
with $w=1+\sqrt{2}$. These boundaries, which are a generalization of Eq.(\ref{eq:GG}), can be seen in Fig.~\ref{fig:pocketDeph}. 

\begin{figure}[ht!]
\centerline{\includegraphics[width=0.3\textwidth]{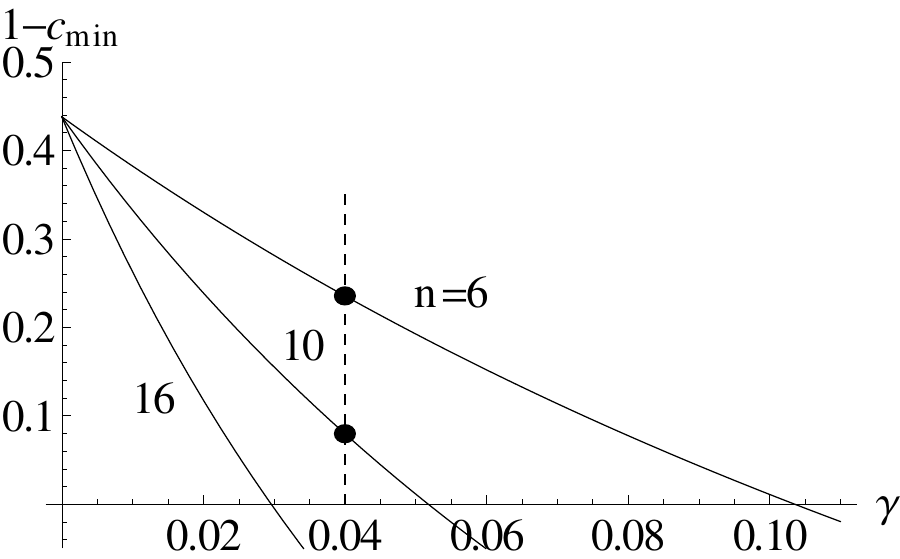}
\includegraphics[width=0.2\textwidth]{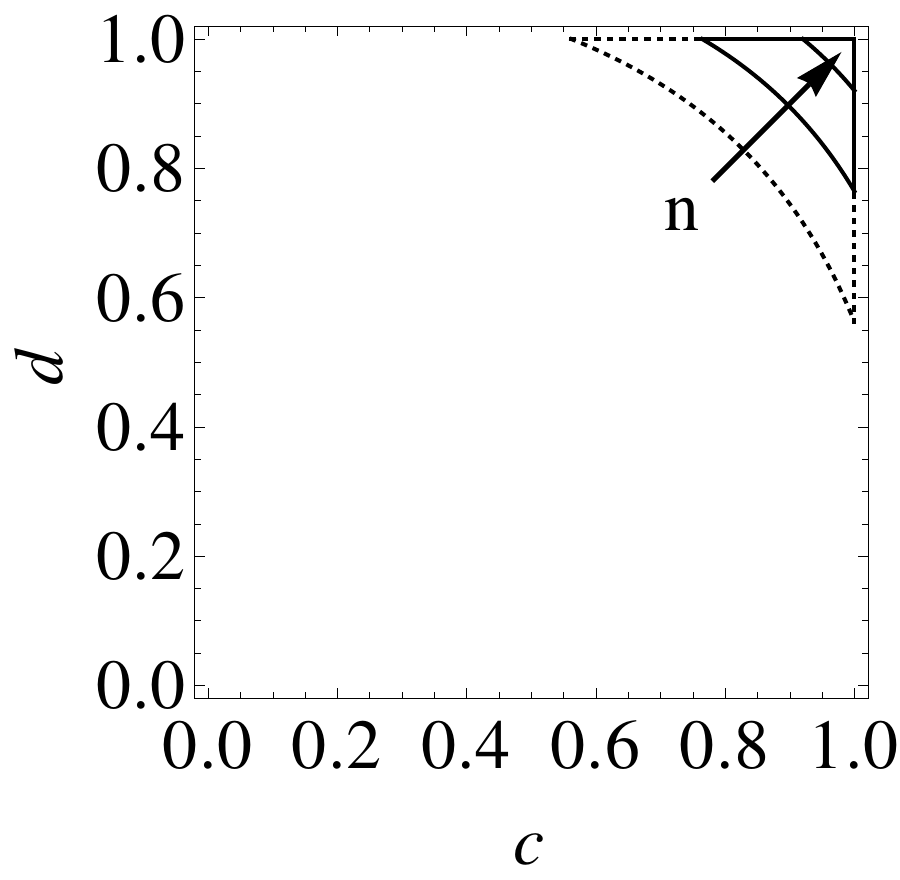}}
\caption{Left: Dependence of size of entangled region, $1-c_{\rm min}$, on dephasing $\gamma$ for three different chain sizes $n=6, 10, 16$. One can see that for $\gamma > \gamma_{\rm c}$ there are no entangled regions ($c_{\rm min}$ gets larger than $1$). Right: The extent of the entangled region for $\gamma=0.04$ and $n=6$ and $n=10$ (two full curves). $c_{\rm min}$ for these two curves is marked with points in the left plot. For reference we also plot the entangled region without dephasing, $\gamma=0$ (dashed curve; the same data as in Fig.~\ref{fig:bulk}). The arrow shows the direction in which $n$ increases; only one quadrant of the ``$c-d$'' plane is shown. Both plots are for $\Gamma=1$ and $\rho_{n/2,n/2+1}$.}
\label{fig:dephG}
\end{figure}

When one increases $\gamma$ from zero the last entangled NESS to survive at $\gamma_{\rm c}$ is the one with $c=d=1$, i.e., $\mu=2$ and $\bar{\mu}=0$. At fixed $\gamma$ and in the limit $n \to \infty$ the critical dephasing strength goes to zero and entanglement disappears! With $\gamma$ there is therefore a discontinuous transition from entangled NESSs at $\gamma=0$ (red pocket in Fig.~\ref{fig:smile}) to no entanglement for $\gamma >0$. This is illustrated in Fig.~\ref{fig:dephG}. One can see how the size of the parameter region with entangled NESSs decreases when one increases $\gamma$ or $n$.
If nearest-neighbor spins $\rho_{j,j+1}$ are in the bulk but not in the middle of chain, $j \neq n/2$, the situation is similar as for $\rho_{n/2,n/2+1}$. As one increases dephasing $\gamma$ or $n$ there is a transition from entangled to separable $\rho_{j,j+1}$ at certain critical dephasing. The only difference is that this critical dephasing weakly depends on the location $j$ in the chain where the two spins are. Critical $\gamma_{\rm c}$ weakly increases as one moves away from the center of the chain. 

For nearest-neighbor spins at the boundary, described by the reduced density matrix $\rho_{12}$, entanglement is possible only if $\Gamma$ is larger than some minimal value $\Gamma_{\rm m}$. This minimal value depends on $\gamma$ and $n$ in a complicated way, however, it has a very simple thermodynamic limit. Namely, $\lim_{n \to \infty} \Gamma_{\rm m}=1$ irrespective of dephasing strength $\gamma$. For $\Gamma>1$ dephasing therefore does not destroy entanglement in $\rho_{12}$. The region of entangled states still has the same shape as in zero dephasing, Fig.~\ref{fig:bG2}, containing all states with $c=1$, only its width $c_{\rm min}$ changes. It shrinks to zero in the limit $n \to \infty$ as well as for $\gamma \to \infty$. In the presence of dephasing there is no entanglement between non-nearest-neighbor spins.

\section{Conclusion}
We have studied bipartite entanglement between two spins in a nonequilibrium stationary state of the XX chain of any length. In the absence of dephasing we show that the two boundary spins can become entangled for arbitrarily small driving away from an infinite temperature equilibrium state. In the bulk, driving must surpass a certain value in order to have nearest-neighbor entanglement. Coupling with the reservoirs also has to be sufficiently strong. Dephasing on the other hand decreases entanglement. Apart from the two boundary spins, entanglement is present only if a product of dephasing strength and system size is sufficiently small.


\begin{thebibliography}{1}
\bibitem{aba} In many works small systems are studied in which thermodynamic quantities, like temperature, are ill-defined.
\end{thebibliography}
\end{document}